\documentclass[aps,prd,twocolumn,a4paper,superscriptaddress,floatfix]{revtex4-1}
\usepackage{graphicx,color,amsmath}
\begin{document}

\newcommand{\be}{\begin{equation}}
\newcommand{\ee}{\end{equation}}
\newcommand{\bq}{\begin{eqnarray}}
\newcommand{\eq}{\end{eqnarray}}
\newcommand{\bw}{\begin{widetext}}
\newcommand{\ew}{\end{widetext}}
\newcommand{\ba}{\begin{align}}
\newcommand{\ea}{\end{align}}
\newcommand{\bc}{\begin{center}}
\newcommand{\ec}{\end{center}}

\title{Scaling solutions of wiggly cosmic strings}
\author{A. R. R. Almeida}
\email{Ana.Almeida@astro.up.pt}
\affiliation{Centro de Astrof\'{\i}sica da Universidade do Porto, Rua das Estrelas, 4150-762 Porto, Portugal}
\affiliation{Instituto de Astrof\'{\i}sica e Ci\^encias do Espa\c co, CAUP, Rua das Estrelas, 4150-762 Porto, Portugal}
\affiliation{Faculdade de Ci\^encias, Universidade do Porto, Rua do Campo Alegre 687, 4169-007 Porto, Portugal}
\author{C. J. A. P. Martins}
\email{Carlos.Martins@astro.up.pt}
\affiliation{Centro de Astrof\'{\i}sica da Universidade do Porto, Rua das Estrelas, 4150-762 Porto, Portugal}
\affiliation{Instituto de Astrof\'{\i}sica e Ci\^encias do Espa\c co, CAUP, Rua das Estrelas, 4150-762 Porto, Portugal}

\date{26 May 2021}

\begin{abstract}
Cosmic strings may have formed in the early universe due to the Kibble mechanism. While string networks are usually modeled as being of Nambu-Goto type, this description is understood to be a convenient approximation, which neglects the typically expected presence of additional degrees of freedom on the string worldsheet. Previous simulations of cosmic strings in expanding universes have established beyond doubt the existence of a significant amount of short-wavelength propagation modes (commonly called wiggles) on the strings, and a wiggly string extension of the canonical velocity-dependent one-scale model has been recently developed. Here we improve the physical interpretation of this model, by studying the possible asymptotic scaling solutions of this model, and in particular how they are affected by the expansion of the universe and the available energy loss or transfer mechanisms---e.g., the production of loops and wiggles. In addition to the Nambu-Goto solution, to which the wiggly model reduces in the appropriate limit, we find that there are also solutions where the amount of wiggliness can grow as the network evolves or, for specific expansion rates, become a constant. Our results show that full scaling of the network, including the wiggliness, is much more likely in the matter era than in the radiation era, which is in agreement with numerical simulation results.
\end{abstract}
\maketitle

\section{\label{sint}Introduction}

Topological defects are ubiquitous in both condensed matter and cosmological contexts. In the latter context they form due to the Kibble mechanism, and the most interesting and well-studied example are cosmic strings \cite{Kibble76}. Their nonlinear nature and non-trivial interactions imply that the detailed quantitative understanding of their properties and observational consequences is a significant challenge \cite{VSH}.

One standard approach is based on Nambu-Goto or Abelian-Higgs (field theory) numerical simulations \cite{BB,AS,ABELIAN,FRAC,RSB,VVO,Stuckey,Blanco,Hiramatsu,Correia1,Correia2}. These are technically difficult and computationally costly, although a recently optimized and validated GPU-based coupling has removed (or at least mitigates) the latter bottleneck \cite{GPU1,GPU2}. A complementary analytic approach is based on the notion of using the knowledge of the \textit{statistical physics} of the string network to obtain a description of its \textit{thermodynamics}. In the simplest case of Nambu-Goto string networks, which have been the subject of most studies so far, the velocity-dependent one-scale (VOS) model \cite{MS2,MS3,MS4,Book,Correia1} has been exhaustively studied, and its quantitative success in describing the large-scale features of the network (including the existence of attractor scaling solutions and the behaviour in the radiation to matter transition) has been demonstrated by direct comparison with both Nambu-Goto and Abelian-Higgs numerical simulations \cite{FRAC,ABELIAN,Correia1,Correia2}. The model allows one to describe the scaling laws and large-scale properties of string networks in both cosmological and condensed matter settings with a minimal number of free parameters. We emphasize that this model---or indeed any other thermodynamical model---needs to be calibrated against numerical simulations, and therefore any such model can only be as accurate as the numerical simulations available.

Cosmologically realistic string networks are not expected to be of Nambu-Goto type---at most, this description is a simple and convenient approximation. More elaborate approaches have been introduced with the goal of explicitly describing the behavior of small-scale structures that are known to be part of realistic cosmic strings \cite{ACK,POLR}. In particular, the previously mentioned simulations of cosmic strings in expanding universes have established beyond doubt the existence of a significant amount of short-wavelength propagation modes (commonly called \textit{wiggles}) on the strings, on scales that can be several orders of magnitude smaller than the correlation length. In previous work \cite{PAP1,PAP2} a mathematical formalism suitable for the description of the evolution of both large-scale and small-scale properties of a cosmic string network in expanding space has been introduced. The wiggly VOS model equations were first obtained in \cite{PAP1}, which only studied two particular limits thereof, the tensionless limit and the linearized limit. Then \cite{PAP2} studied the particular case where the wiggliness was assumed to reach scaling, and also made a brief comparison with the simulations of \cite{FRAC}. The necessary formalism has been recently extended to the case of generic current-carrying cosmic strings \cite{Currents}.

In this work we continue the exploration of the properties of wiggly cosmic strings, by improving its physical interpretation and modelling. Specifically, we study the effect of two key dynamical mechanisms impacting the dynamics of cosmic string networks---the expansion of the universe, and energy loss or transfer mechanisms (e.g., the production of loops and wiggles)---on the various possible scaling solutions for these networks. This enables a systematic study of the full set of asymptotic scaling solutions; as we demonstrate in what follows, one may find solutions with wiggliness growing, disappearing, or reaching scaling. The balance of these dynamical mechanisms, quantified by the VOS model parameters, determines the conditions under which each of the solutions applies.

Our long-term goal is to have an accurately calibrated wiggly VOS model that reproduces the key dynamical features of numerical simulations. For the standard (non-wiggly) VOS model, the feasibility of this approach has recently been demonstrated \cite{Correia1,Correia2}, provided one has hundreds to thousands of high-resolution simulations, covering tens of different expansion rates---which is itself feasible with a fast and efficient GPU-based code \cite{GPU1,GPU2}. However, for the wiggly model the same approach is currently not possible, for the simple reason that current simulations do not measure and output the key wiggliness diagnostics (e.g., the renormalized mass per unit length, which is an explicit parameter in the wiggly model, or the multifractal dimension). The one exception to this is \cite{FRAC}, which published some of these diagnostics from Nambu-Goto simulations. However, these simulations had relatively low resolution (by present day standards) and were only done for three expansion rates (matter and radiation era, plus Minkowski space), and it has been previously shown that these are not sufficient for a detailed calibration \cite{PAP2}.

Our goal in the present work is to set the stage for a future detailed calibration of this wiggly VOS model. By characterizing all possible scaling solutions, and particularly the behaviour of wiggliness therein (i.e., under what conditions wiggliness grows, disappears, or reaches scaling) we can plan a subsequent programme of field theory and Nambu-Goto simulations optimized for this calibration, which can test the consistency of the model solutions by carrying out simulations with various different expansion rates and including suitable numerical diagnostics for the wiggliness of the strings.

We start with a brief comparative study of Nambu-Goto and wiggly cosmic strings in Sect. \ref{sevo}, with the goal of making the present work self-contained. The possible asymptotic scaling solutions without and with cosmological expansion are then presented, respectively in Sect. \ref{flat} and Sect. \ref{expand}. Broadly speaking, we find three classes of solution. The first is the trivial Nambu-Goto one, to which, as expected, the wiggly model reduced to in the appropriate physical limit. The two other model classes are new: the amount of wiggliness can either grow as the network evolves or, under relatively specific conditions, become a constant. In particular we discuss how the existence (or not) of these solutions depends on the phenomenological parameters of the VOS model. Finally we present some conclusions and an outlook in Sect. \ref{concl}.

\section{\label{sevo}Nambu-Goto and wiggly strings}

Here we present a brief review of the mathematical formalism behind the VOS model for Nambu-Goto strings, in order to introduce the relevant notation; a more thorough discussion of the model can be found in \cite{Book}. We then summarize how the model is extended to wiggly case, motivating the evolution equations that will be the subject of the exploration of the present work; their detailed derivation can be found in \cite{PAP1,PAP2}.

\subsection{Nambu-Goto strings}

The first assumption in the VOS approach is to \textit{localise} the strings so that we can treat them as one-dimensional line-like objects. This is clearly a good assumption for local strings, but we note that it has been shown to work well even for strings possessing long-range interactions \cite{MS4}.

The second step is to \textit{average} the microscopic string equations of motion to derive the key evolution equations for the average root mean squared (RMS) string velocity $v$ and characteristic length scale $L$. This is a generalisation of Kibble's original one-scale model \cite{KIB,BMOD}, and has been described in detail elsewhere \cite{MS2,MS3,Book}. Kibble's model describes string motion in terms of a single \textit{characteristic length scale}, denoted $L$. In particular, this assumption implies that this length scale coincides with the string \textit{correlation length} $\xi$ and the string \textit{curvature radius} $R$. We stress that this is an approximation which can be tested numerically \cite{FRAC,ABELIAN}. By incorporating a variable RMS velocity $v$, the VOS model significantly extends its validity, including into regimes with frictional damping and across the important matter-radiation transition, thus giving a quantitative picture of the complete history of a cosmic string network.

Specifically one starts by defining the string network energy $E$ and RMS velocity $v$
\bq
E&=&\mu_0 a(\tau)\int\epsilon d\sigma \\
v^2&=&\frac{\int{\dot{\bf x}}^2\epsilon d\sigma}{\int\epsilon d\sigma}\,,
\label{eee}
\eq
where $\mu_0$ is the string mass per unit length. String networks divide fairly neatly into two distinct populations, \textit{viz.} long (or `infinite') strings and small closed loops. The long string network is, to a good approximation \cite{FRAC}, a Brownian random walk on large scales and can be characterized by a characteristic length scale (or inter-string separation) $L$, which can be used to replace the energy $E= \rho V$ in long strings
\be
\rho \equiv \frac{\mu_0}{L^2}\,.
\ee
A phenomenological term necessary to account for the loss of energy from long strings by the production of cosmic string loops, This is described by a \textit{loop chopping efficiency} parameter $c$ such that
\begin{equation}
\left(\frac{d\rho}{dt}\right)_{\rm to\ loops}= c v\frac{\rho}{L}
\, . \label{rtl}
\end{equation}
In this one-scale approximation, we would expect the loop parameter $c$ to remain constant irrespective of the cosmic regime, because it is multiplied by factors which determine the string network self-interaction rate.

From the microscopic string equations of motion, one can then average to derive the evolution equation for the length scale $L$ and the velocity $v$
\bq
2\frac{dL}{dt}&=&2HL(1 + {v^2})+c v \\ \label{evl0}
\frac{dv}{dt}&=&\left(1-{v^2}\right)\left[\frac{k(v)}{L}-2Hv\right]\,, \label{evv0}
\eq
where $H$ is the Hubble parameter and where $k$ is called the \textit{momentum parameter}, which at a phenomenological level is expected to include the effects of small-scale wiggles (more on this in what follows). Note that strictly speaking it would be the curvature radius $R$ which should appear in the denominator of the first term of the velocity equation, but the one-scale assumption implies that we can identify $R=\xi=L$; nevertheless one should keep this distinction in mind in more general situations \cite{ABELIAN}. Detailed descriptions of the behaviour of the parameter $k$ can be found in \cite{MS3,Correia1}.

It is well known \cite{Book} that attractor scale-invariant solutions with $L\propto t\propto H^{-1}$ and $v=const.$, only exist when the scale factor is a power law of the form $a(t)\propto t^\lambda$. Introducing the convenient parameter $L=\zeta t$ this solution can be written in the following implicit form
\bq
\zeta^2_{NG}&=&\frac{k(k+c)}{4\lambda(1-\lambda)}\\ v^2_{NG}&=&\frac{k(1-\lambda)}{\lambda(k+c)}\,.
\label{scalsol}
\eq
Note that since the velocity is a constant in this solution and $k(v)$ is a function of velocity, $k$ itself becomes a constant in this limit. Therefore, $k$ in this solution denotes the constant value of $k(v)$ given by solving the second (implicit) equation for the velocity. It's easy to verify numerically that this solution is well-behaved and stable for all realistic parameter values. 

If the scale factor is not a power law, then simple scale-invariant solutions like (\ref{scalsol}) do not exist. Physically this happens because the network dynamics is unable to adapt rapidly enough to the changes in the background cosmology. Examples of this are the transition between the radiation and matter-dominated eras and the onset of dark energy domination around the present day \cite{Azevedo}. Indeed, since this relaxation to a changing expansion rate is rather slow, strictly speaking realistic cosmic string networks are \textit{never} in scaling during the matter-dominated era \cite{MS3,FRAC}.

\subsection{\label{wigmic}Wiggly strings}

The above framework can be extended to more general models, building upon earlier works of Carter \cite{CARTERA,CARTERB}. For what follows the key difference is that wiggly strings have an energy density in the locally preferred string rest frame (denoted $U$) and a local string tension (denoted $T$) which are not identical constants as in the Nambu-Goto case, but have different dependencies on a dimensionless parameter $w$,
\bq
T&=&w\mu_0\\ U&=&\frac{\mu_0}{w}\, ,\label{wigt}
\eq
so that $T/U=w^2$.

Now the total energy of a piece of string is
\begin{equation}
E=a\int\epsilon Ud\sigma=\mu_0 a\int\frac{\epsilon}{w}d\sigma\, .\label{enertot}
\end{equation}
Part of this is the bare energy that can be ascribed to the string itself,
\begin{equation}
E_0=\mu_0 a\int\epsilon d\sigma\,, \label{enerst}
\end{equation}
while the rest is in the small-scale wiggles
\begin{equation}
E_w=\mu_0 a\int\frac{1-w}{w}\epsilon d\sigma\,.\label{enerwg}
\end{equation}
Each of these energies can yield a characteristic length scale for the string network. The total length should be the length that a Nambu-Goto string with the same total energy would have, while the bare correlation length measures the characteristic length of a Brownian network. Specifically, the string correlation length is defined with respect to the bare string density
\begin{equation}
\rho_0\equiv\frac{\mu_0}{\xi^2} \,, \label{wig_b2}
\end{equation}
and we assume that the correlation length thus defined is approximately equal to the string curvature radius
\begin{equation}
\xi\sim R \, ; \label{wig_b3}
\end{equation}
such an assumption can be tested numerically \cite{FRAC}, and although not exact is sufficiently accurate for our present purposes. The correlation length $\xi$ still has a physically clear meaning, while the characteristic length scale $L$ is now only a proxy for the total energy in the network,
\begin{equation}
\rho\equiv\frac{\mu_0}{L^2} \, . \label{wig_b2newL}
\end{equation}
Note that this means that
\begin{equation}
\xi^2=\mu L^2 \,,\label{ximuL}
\end{equation}
where $\mu$ is the renormalized string mass per unit length $\mu$, defined as
\begin{equation}
\mu\equiv\frac{E}{E_0}=\langle w\rangle^{-1}=\langle w^{-1}\rangle_0\, . \label{vwwv}
\end{equation}
Strictly speaking, $\mu$ is a scale-dependent quantity, $\mu=\mu(\ell,t)$ \cite{FRAC}, but the $\mu$ thus defined is to be understood as a quantity measured at a mesoscopic scale somewhat smaller than the horizon. Intuitively, an obvious choice might be the coherence length in the model itself, but this is not mandatory.

Physically, the natural way to include small-scale structure in the analytic model is through an evolution equation for $\mu$, and this is the approach followed in the wiggly VOS model. Alternatively we could define a characteristic scale for the energy in the wiggles: defining $\rho_w=\mu_0/S_w^2$ and noting that $\rho_w=\rho-\rho_0$ we have
\begin{equation}
S_w=\frac{\xi}{\sqrt{\mu-1}}=\sqrt{\frac{\mu}{\mu-1}}L\,.\label{scalewig}
\end{equation}
An approach along these lines is that of the three-scale model \cite{ACK}.

From the point of view of our analytic modeling, we are no longer allowed to identify the three natural length scales we considered in the Nambu-Goto case. In other words, we can no longer have a one-scale model. Therefore, an averaged model for wiggly cosmic string evolution contains three (rather that two) evolution equations. Apart from evolution equations for a length scale and velocity, one needs a third equation which describes the evolution of small-scale structure. This is reminiscent of the three-scale model \cite{ACK}, with two crucial differences. First, in the three scale model all three evolution equations do in fact describe length scales, while in our case only one of them does so (although a second equation can dependently be converted into one that does). Second, in the three scale model there is no allowance for the evolution of the string velocities.

One must also rethink the definition of the averages. Specifically, when one is defining average quantities over the string network (say, the average RMS velocity), should the average be over the total energy
\begin{equation}
\langle {\dot{\bf x}}^2 \rangle=\frac{\int{\dot{\bf x}}^2U\epsilon d\sigma}{\int U\epsilon d\sigma}=\frac{\int Q\frac{\epsilon}{w}d\sigma}{\int \frac{\epsilon}{w}d\sigma}\, ,\label{sexta1}
\end{equation}
or just the energy in string
\begin{equation}
\langle {\dot{\bf x}}^2 \rangle_0=\frac{\int
{\dot{\bf x}}^2\epsilon d\sigma}{\int\epsilon d\sigma}\, ? \label{sexta2}
\end{equation}
In other words, should pieces of string that have larger mass currents be given more weight in the average? Given the discussion so far, the first definition is more natural (and it is the one adopted in what follows), although the alternative choice deserves further discussion---see \cite{Currents} for an alternative approach. These two different averaging procedures can be applied to any other relevant quantity. For a generic quantity $Q$, the two averaging methods are related via
\begin{equation}
\langle Q \rangle=\frac{\langle QU \rangle_0}{\langle U \rangle_0}=\frac{\langle Q/w\rangle_0}{\mu}\,.\label{twoaverages}
\end{equation}

One also needs phenomenological terms describing how energy is exchanged between the bare string and the wiggles. In the original Nambu-Goto VOS model, long string intercommutings did not affect the evolution of the network and so we did not need to directly model them. For wiggly strings one must consider these intercommutings, since they increase the number of kinks on the string network---and consequently add energy to the wiggles.

By analogy with Eq. (\ref{rtl}), we define the fraction of the bare energy density lost into loops per unit time as
\begin{equation}
\left(\frac{1}{\rho_0}\frac{d\rho_0}{dt}\right)_{\rm loops}=-c f_0(\mu) \frac{v}{\xi}\, . \label{wig_b4}
\end{equation}
Numerical simulations suggest that small-scale structure might enhance loop production, and we phenomenologically allow for this possible enhancement by allowing for an explicit dependence on $\mu$, encoded in a function $f_0(\mu)$ which becomes unity in the Nambu-Goto limit. Moreover, whenever two strings inter-commute, kinks are produced whether or not loop production occurs. This corresponds to energy being transferred from the bare string to the small-scale wiggles, which we model as
\begin{equation}
\left(\frac{1}{\rho_0}\frac{d\rho_0}{dt}\right)_{\rm wiggles}= -cs(\mu)\frac{v}{\xi}\, , \label{wig_b6}
\end{equation}
in analogy with the previous term for losses into loops. Here $s$ vanishes in the Nambu-Goto limit, and it should include the effects of kink decay on long strings due to gravitational radiation.

As for the fraction of the total energy lost into loops, we need to take into account that the energy may be come from the bare string or from the wiggles
\begin{equation}
\left(\frac{1}{\rho}\frac{d\rho}{dt}\right)_{\rm loops}=\left(\frac{1}{\rho}\frac{d\rho_0}{dt}\right)_{\rm loops}+\left(\frac{1}{\rho}\frac{d\rho_w}{dt}\right)_{\rm loops}\,. \label{wig_b5a}
\end{equation}
The energy loss from the bare string has already been characterized by the parameter $f_0$ in Eq. (\ref{wig_b4}). Defining an analogous term for the losses from the wiggles
\begin{equation}
\left(\frac{1}{\rho_w}\frac{d\rho_w}{dt}\right)_{\rm loops}=-c f_1(\mu) \frac{v}{\xi}\,, \label{wig_b5b}
\end{equation}
we end up with
\begin{equation}
\left(\frac{1}{\rho}\frac{d\rho}{dt}\right)_{\rm loops}=-c\left[\frac{f_0}{\mu}+f_1\left(1-\frac{1}{\mu}\right)\right]\frac{v}{\xi}=-c f(\mu) \frac{v}{\xi}\,, \label{wig_b5}
\end{equation}
where we defined an overall loss parameter $f$ which may also have a dependence on $\mu$, to account for the fact that loops are preferentially produced from regions of the long string network containing more small-scale structure than average. There is evidence of this fact from numerical simulations \cite{BB,AS,FRAC}. Somewhat similar parameters have been introduced before \cite{ACK}, but those are typically constant and defined as the excess kinkiness of a loop compared to a piece of long string of the same size. In what follows we will use the following simplifying assumptions, previously discussed in \cite{PAP2}
\be\label{enloss}
\begin{split}
f_0(\mu)&=1\\
f(\mu)&=1+\eta\left(1-\frac{1}{\sqrt{\mu}}\right)\\ s(\mu)&=D\left(1-\frac{1}{\mu^2}\right)\,.
\end{split}
\ee
Here we have introduced two additional phenomenological parameters, $\eta$ and $D$, which are specific to the wiggly VOS model. (In particular $\eta$ should not be confused with conformal time.)

Finally, we note the renormalized string mass per unit length is defined at a renormalization scale $\ell$ that need not be constant---for example, one could choose this scale to be that of the correlation length, which will be time dependent. Changing this scale is tantamount to redefining what small-scale structure is, and thus must have an effect on the value of $E_{0}$ (as well as $v$ since $w$ will be correspondingly changed). This is accounted for by introducing the following scale-drift terms
\begin{equation}
\frac{1}{\mu}\frac{\partial\mu}{\partial\ell}\frac{d\ell}{dt}\sim\frac{d_{m}-1}{\ell}\frac{d\ell}{dt}\label{dm}
\end{equation}
\begin{equation}
\frac{\partial v^{2}}{\partial\ell}\frac{d\ell}{dt}=\frac{1-v^{2}}{1+\left\langle w^{2}\right\rangle}\frac{\partial\left\langle w^{2}\right\rangle}{\partial\ell}\frac{d\ell}{dt}\label{v_dm}
\end{equation}
where $d_{m}\left(\ell\right)$ is the multifractal dimension of a string segment at scale $\ell$ \cite{TAKAYASU}. Note that Eq.~(\ref{dm}) is essentially just a geometric identity whereas Eq.~(\ref{v_dm}) comes from imposing total energy conservation across different scales.

With these definitions, and further assuming uniform wiggliness (i.e., $w$ to be just a function of time, at least locally) one can obtain the following system of equations, derived in detail in \cite{PAP1,PAP2}
\begin{widetext}
\begin{equation}
2\frac{dL}{dt}=HL\left[3+v^2-\frac{(1-v^2)}{\mu^2}\right]+\frac{cfv}{\sqrt{\mu}} \label{ell_full}
\end{equation}
\begin{equation}
2\frac{d\xi}{dt}=H\xi\left[2+\left(1+\frac{1}{\mu^{2}}\right)v^{2}\right]+v\left[k\left(1-\frac{1}{\mu^{2}}\right)+c\left(f_{0}+s\right)\right]
+\left[d_{m}\left(\ell\right)-1\right]\frac{\xi}{\ell}\frac{d\ell}{dt}\label{qui_full}
\end{equation}
\begin{equation}
\frac{dv}{dt}=\left(1-v^{2}\right)\left[\frac{k(v)}{\xi\mu^{2}}-Hv\left(1+\frac{1}{\mu^{2}}\right)-\frac{1}{1+\mu^{2}}\frac{\left[d_{m}\left(\ell\right)-1\right]}{v\ell}\frac{d\ell}{dt}\right]\label{v_full}
\end{equation}
\begin{equation}
\frac{1}{\mu}\frac{d\mu}{dt}=\frac{v}{\xi}\left[k\left(1-\frac{1}{\mu^{2}}\right)-c\left(f-f_{0}-s\right)\right]-H\left(1-\frac{1}{\mu^{2}}\right)
+\frac{\left[d_{m}\left(\ell\right)-1\right]}{\ell}\frac{d\ell}{dt}\label{mu_full}
\end{equation}
\end{widetext}
where as usual $H\equiv\dot{a}/a$ is the Hubble parameter. Note that the first, second and fourth equations are not independent, being related by Eq. (\ref{ximuL}). In this work we restrict ourselves to solutions with a constant renormalization scale (leaving the case with a running scale for subsequent work), so henceforth we set $d\ell/dt=0$. 

The key question in cosmic string evolution is whether the small-scale component also scales, i.e., whether we should also expect $\mu$ to evolve towards a constant value. Despite current simulations not answering this question definitely \cite{FRAC}, they suggest that such a small-scale scaling is reached at least in a matter era (when $\lambda=2/3$). In the radiation era simulations show a more complex behavior, which could reflect the fact that the approach to scaling is slower in this case (since there is less Hubble damping) or could be due to the absence of such a solution. In what follows we will shed light on this issue.

Clearly, the dynamical equations for wiggly model include two different physical mechanisms: expansion and energy losses. We will study the effects of each of them on the possible asymptotic scaling solutions. In general we will assume
\be\begin{split}
    L &= \zeta_0 t^\alpha\\
    v &= v_0t^\beta\\
    \mu &= m_0t^\gamma\,.
\end{split}\ee
Recall that $\xi^2=\mu L^2$, which also leads to
\be
\xi=\sqrt{m_0}\zeta_0 t^{\alpha+\gamma/2}\,;
\ee
finally, when studying solutions in the expanding universe we will assume the aforementioned scale factors, with $a\propto t^\lambda$. We also note that all these asymptotic scaling solutions will either have $v=const.$ or $v\longrightarrow0$, and that the function $k(v)$ becomes a constant in both limits. Thus in the sections hat follow we will simply write $k$ when discussing each of the solutions, but it is important to bear in mind that this denotes the constant value of $k(v)$ in each case.

\section{\label{flat}Scaling solutions without expansion}

We start by discussing the case without expansion, thus setting $H=0$. In this case we expect the VOS model to apply for $k=0$ \cite{MS4,FRAC,Book,Correia1}. We also note that linear scaling has been numerically seen in Minkowski space \cite{Sakellariadou,FRAC}, and we will presently show that this is indeed a possible solution of the wiggly model. In what follows we separately treat the cases without and with energy losses. On the other hand we will not discuss the ultra-relativistic $v=1$ solutions; while such solution are mathematically allowed, it is clear that they would only be physically relevant in contracting universes \cite{Cabral}.

\subsection{Without energy losses}

In this case we have a trivial solution, since there is no dynamical mechanism affecting the string network: all three dynamical quantities on the VOS model must necessarily be constant
\be\label{case1}
\begin{split}
    L &= \zeta_0\\
    v &= v_0\\
    \mu &= m_0\\
    \xi &= \sqrt{m_0}\zeta_0\,,
\end{split}
\ee
with the constant values being undetermined. In practice, they could be determined by measuring them in a numerical simulation, as a means to check whether any choice of values of the parameters is allowed.

The physical interpretation of this solution is equally simple. In the absence of energy loss mechanisms, we have a string network in an equilibrium configuration, with a constant energy density and velocity, and also with a constant wiggliness.

\subsection{With energy losses}

Here we assume that the energy loss terms are of the form given by Eq. (\ref{enloss}). The presence of the energy losses leads to a different behaviour for the characteristic length scale $L$, and consequently also for the correlation length $\xi$.

\begin{figure}
\centering
  \includegraphics[width=1.0\columnwidth]{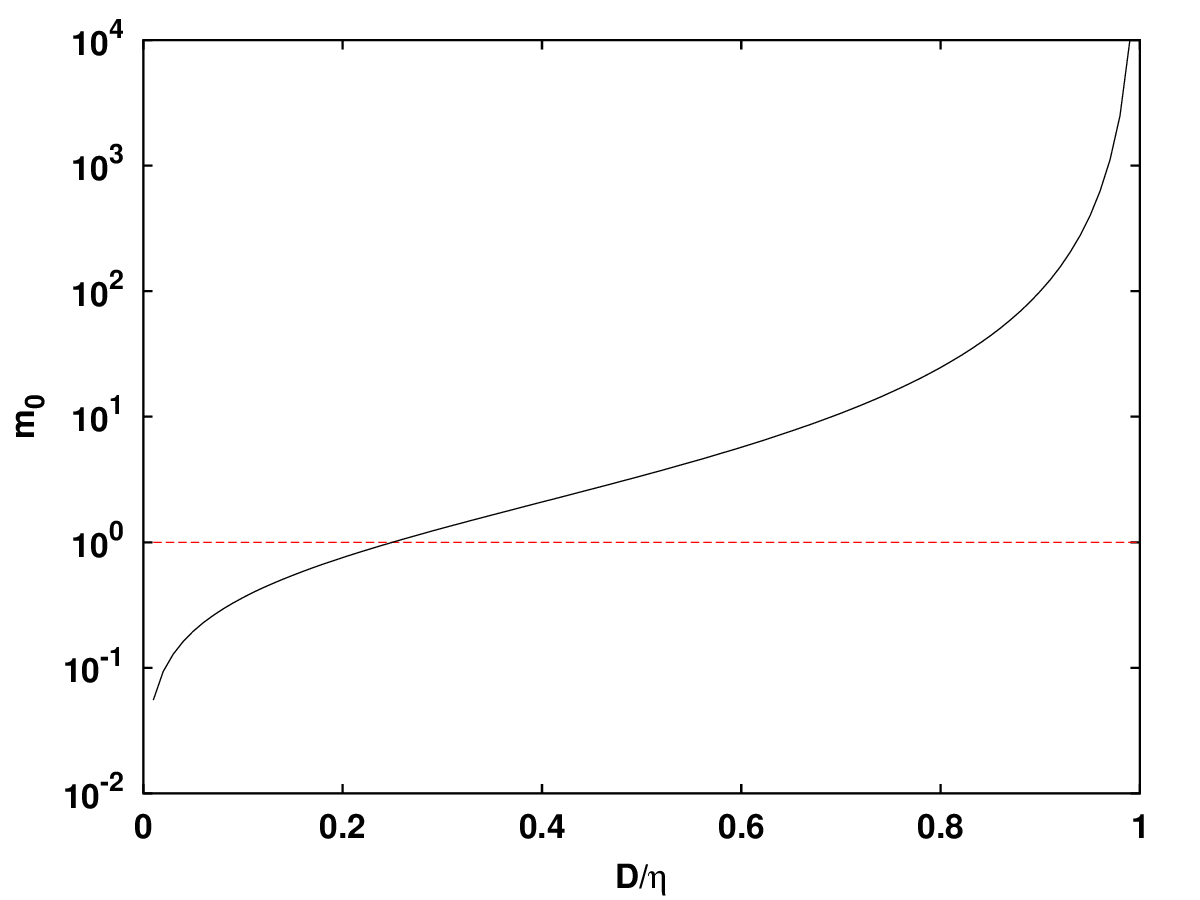}
  \caption{The two real solutions of Eq. (\ref{sol1a}) for the wiggliness $m_0$, as a function of the ratio of the small-scale structure parameters,  $D/\eta$. Note that physically $\mu\ge1$, so a non-trivial $m_0>1$ solution only exists for $1/4<D/\eta<1$.}
  \label{fig01}
\end{figure}

In this case there are two possible regimes. The first one has linear scaling of the characteristic length scale ($\alpha=1$) and  constant velocity and wiggliness ($\beta=\gamma=0$)
\be\label{case2a}
\begin{split}
    L &= \zeta_0 t\\
    v &= v_0\\
    \mu &= m_0\\
    \xi &= \sqrt{m_0}\zeta_0 t\,,
\end{split}
\ee
subject to the consistency constraints
\bq
\zeta_0\sqrt{m_0}&=&\frac{cv_0}{2}\left[1+\eta\left(1-\frac{1}{\sqrt{m_0}}\right)\right]\\ \label{sol1a}
\eta\left(1-\frac{1}{\sqrt{m_0}}\right)&=&D\left(1-\frac{1}{m_0^2}\right)\\ \label{sol1}
\eta&\ge&D\,;
\eq
this includes the simple Nambu-Goto case $m_0=1$ (for $\eta=D$), but larger values of $\mu_0$ are allowed in principle, depending on the values of the two parameters.

For realistic values of the parameters $D$ and $\eta$ (i.e., of order unity and with $\eta\ge D$), Eq, (\ref{sol1a}) has two real and two complex conjugate solutions, and physically realistic solutions for the wiggliness must have $m_0\ge1$. It follows that the allowed solutions are determined by the ratio of the two VOS parameters: if $D/\eta<1/4$ then $m_0=1$ is the only physically allowed solution, while for  $1/4<D/\eta<1$ there is a solution with $m_0>1$, as shown in Fig. \ref{fig01}. In this interpretation, if the amount of small-scale structure produced is too small a fraction of the amount thereof that is removed from the network, then no wiggliness can asymptotically survive on the string network, leading to the trivial Nambu-Goto solution. Conversely, if this ratio is large enough then wiggliness can survive. That said, it is not a priori clear if this $m_0>1$ solution, even if mathematically allowed, is physically realized. One possible argument for this is that if one takes the opposite regime (corresponding to $D>\eta$, as will be discussed presently) and takes the $D\longrightarrow\eta$ limit one finds a $m_0=1$ constant wiggliness solution, and one might therefore expect that this would still be the case as one continues decreasing the $D/\eta$ ratio.

Instead of interpreting $D$ and $\eta$ in terms of the overall amount of energy in wiggles that is produced and removed, one might alternatively interpret them as describing the frequencies \footnote{We thank Jos\'e Pedro Vieira for bringing this interpretation to our attention.}--- specifically, the frequency of events that produce wiggles (e.g., in the form of kinks or cusps) and events that remove wiggliness (e.g., by producing loops). Then a small $D/\eta$ ratio corresponds to the case where any kinks or cusps that are produced are likely to find themselves part of a loop faster than others are produced. We also note that for $m_0\neq1$, and using the definitions of Eq. (\ref{enloss}), Eq. (\ref{sol1a}) can be written in the suggestive form
\be
f=f_0+s\,,
\ee
which becomes trivial for $m_0=1$. This suggests that the solution with $m_0\neq1$ corresponds to a special point in parameter space, where the energy loss terms are precisely balanced, which in turn casts doubt on whether such a solution can be an attractor for the dynamics of the network. Numerical simulations should be needed to clarify these points.

\begin{figure}
\centering
  \includegraphics[width=1.00\columnwidth]{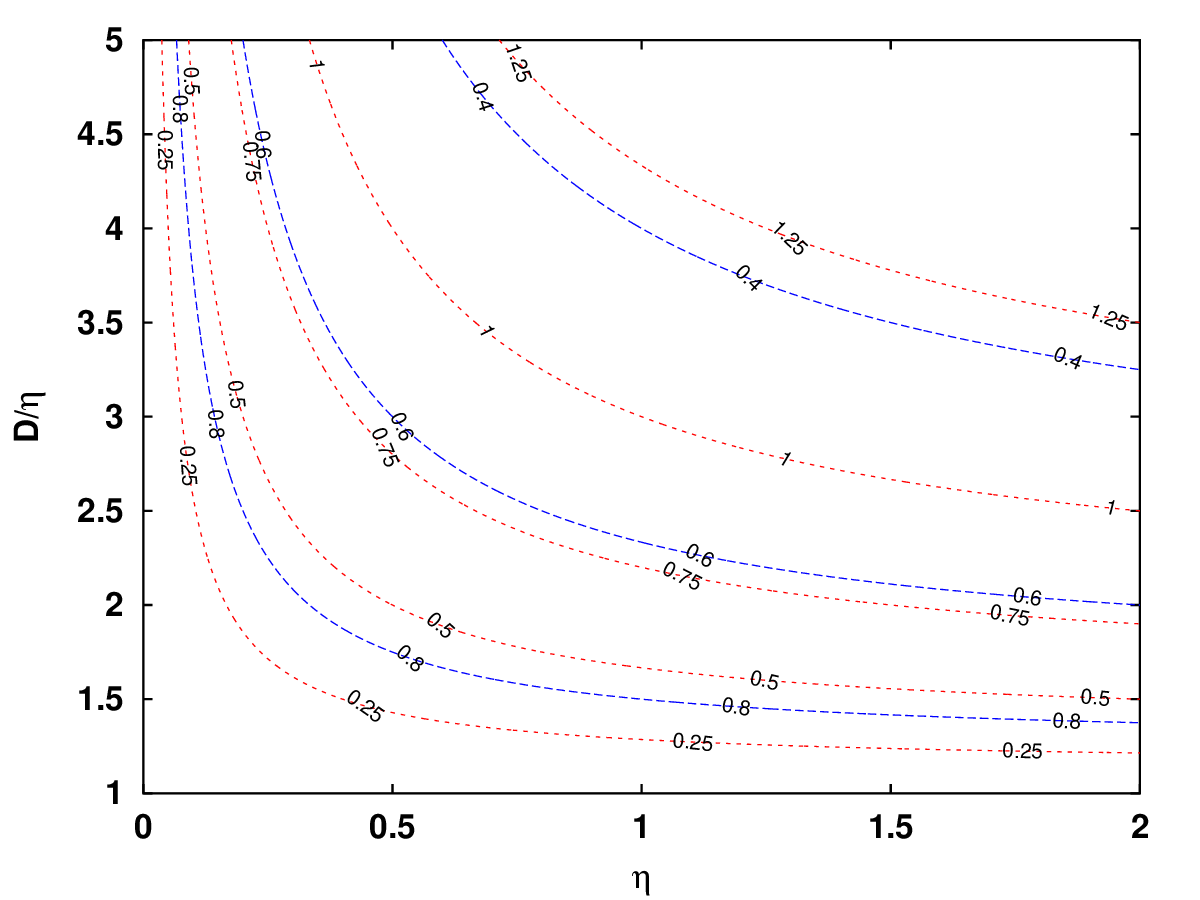}
  \caption{The values of the scaling exponents $\alpha$ (blue dashed contours) and $\gamma$ (red dotted contours), respectively given by Eq. (\ref{expalf}) and  Eq. (\ref{expgam}) for the wiggliness $m_0$, as a function of the small-scale structure parameters $\eta$ and $D/\eta$.}
  \label{fig02}
\end{figure}

The second regime has growing small-scale structure, which is offset by a slower growth of the characteristic length scale (while the velocity remains constant)
\be\label{case2b}
\begin{split}
    L &= \zeta_0 t^\alpha\\
    v &= v_0\\
    \mu &= m_0t^\gamma\\
    \xi &= \sqrt{m_0}\zeta_0 t\,,
\end{split}
\ee
with the following consistency constraints
\bq
\alpha&=&1-\frac{1}{2}\gamma=\frac{1+\eta}{1+D}\\ \label{expalf}
\gamma&=&2\frac{D-\eta}{1+D}\\ \label{expgam}
\zeta_0\sqrt{m_0}&=&\frac{v_0}{2}c(1+D)\\
\eta&<&D\,.
\eq
A visual illustration of the dependence of the scaling exponents $\alpha$ and $\gamma$ on the small-scale structure parameters $\eta$ and $D$ can be found in Fig. \ref{fig02}. Note that, as has already been mentioned, the first solution can be obtained from the second in the limit $D\longrightarrow\eta$ with $m_0\longrightarrow1$. The physical interpretation of this solution is equally clear: $\eta<D$ means that the amount of small scale structure removed from the network (e.g. by loop production) is smaller than the amount produced (e.g. by intercommutings) and therefore the long string wiggliness must grow.

Finally, it is worthy of note these two solutions only differ in the behaviour of the characteristic length $L$ (or equivalently the overall energy density of the network) and the string wiggliness. In both solutions the velocity $v$ is a constant and---more interestingly---the correlation length $\xi$ is always scaling linearly ($\xi\propto t$), regardless of whether or not $L$ does. Noting that $\xi$ is related to the evolution of the bare string energy density, cf. Eq. (\ref{wig_b2}), this shows that in both solutions we have $\rho_0\propto t^{-2}$. On the other hand, the ratio of the total string energy density to the bare one be a constant for the first solution, for which the wiggliness is a constant. In the second solution this ratio grows proportionally to the wiggliness,
\be
\frac{\rho}{\rho_0}=m_0 t^\gamma\,.
\ee
This illustrates the earlier point that we no longer have a one-scale model, and different length scales can have different behaviours.

\section{\label{expand}Scaling solutions in expanding universes}

We now move to the case of expanding universes. Specifically, we assume that the scale factor is a power law of the form $a(t)\propto t^\lambda$, with $0<\lambda<1$. We again ignore the $v=1$ solutions but now also discard the $k=0$ solutions, since in this case we do expect that $k\neq0$, as confirmed by numerical simulations \cite{FRAC}. As in the previous section, we will first consider the case without energy losses and then the general case where they are allowed.

\subsection{Without energy losses}

In this case we find three possible regime. Firstly, we have the canonical Nambu-Goto VOS solution
\be\label{case4a}
\begin{split}
    L &= \frac{k}{2\sqrt{\lambda(1-\lambda)}}t\\
    v &= \sqrt{\lambda^{-1}-1}\\
    \mu &=1\\
    \xi &= \frac{k}{2\sqrt{\lambda(1-\lambda)}}t\,.
\end{split}
\ee
The range of expansion rates for which this is a physically viable solution depends on how physically interprets this velocity. If it is interpreted as a microscopic velocity, then we only require that the string velocity does not exceed the speed of light ($v_0^2<1$), which implies $\lambda>1/2$; in other words, the radiation era is the limiting case. On the other hand, if it is interpreted as an average velocity, then the physical constraint should be $v_0^2\le1/2$ (corresponding to the average velocity of loops in Minkowski space), which leads to $\lambda\ge2/3$, and the matter era is the limiting case. Either way, this confirms the well-known result that in the matter era (but not in the radiation era) damping due to the expansion of the universe is by itself sufficient to ensure the scaling of a Nambu-Goto string network.

A regime with non-trivial constant wiggliness, $\mu_0>1$, also exists, but only for the matter era
\be\label{case4b}
\begin{split}
    L &= \zeta_0t\\
    v &= \frac{1}{\sqrt{1+m_0^2}}\\
    \mu &= m_0\\
    \xi &= \frac{3}{2}kv_0t\,,
\end{split}
\ee
with the consistency relation
\be
\sqrt{m_0}\zeta_0=\frac{3}{2}kv_0\,,
\ee
which is physically viable for any value of $m_0$, and is again commensurate with the indication, from Nambu-Goto numerical simulations \cite{FRAC}, that scaling of small-scale structure is easier to achieve in the matter era than in the radiation era. Note that the choice $m_0=1$ matches the previous Nambu-Goto solution, given by Eq. (\ref{case4a}). Moreover this behaviour is analogous to the one previously identified for chiral superconducting strings \cite{Oliveira}.

\begin{figure}
\centering
  \includegraphics[width=1.00\columnwidth]{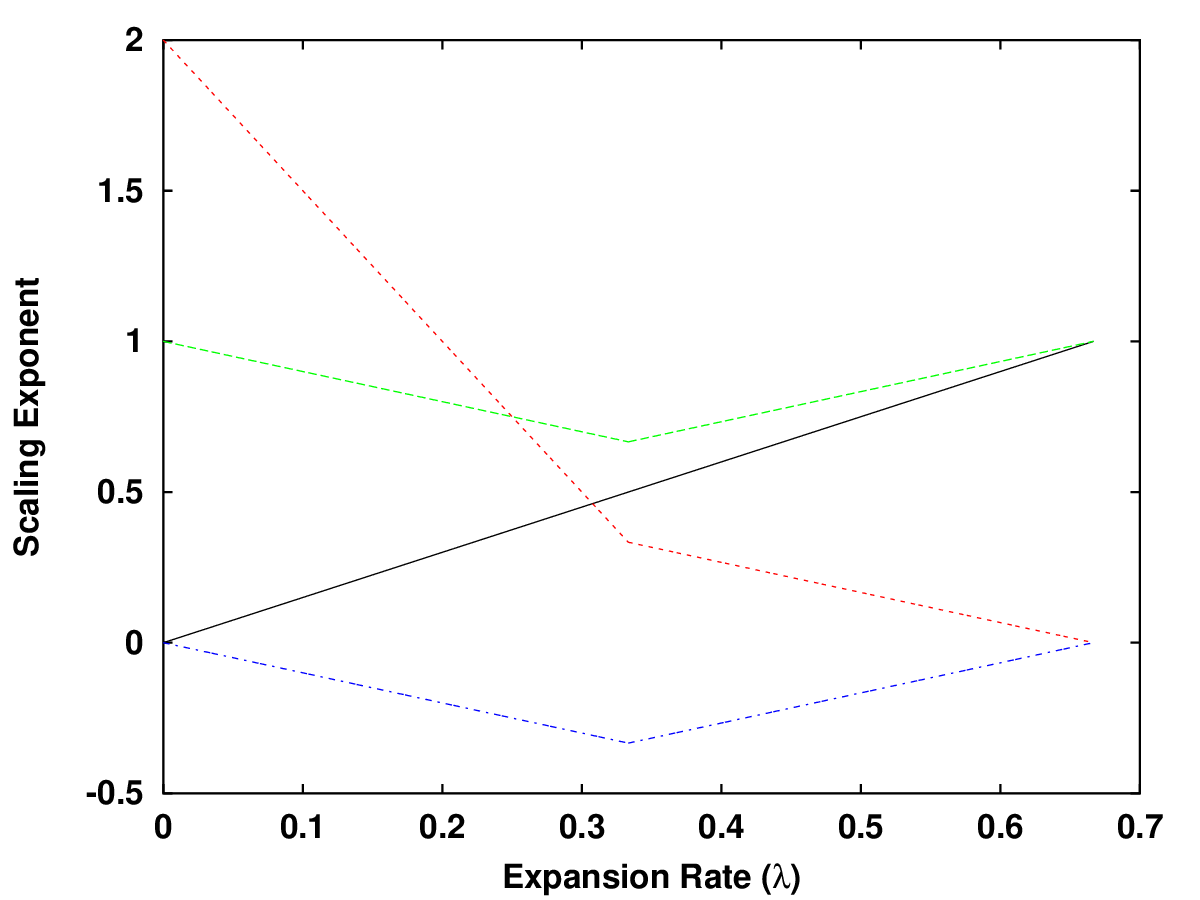}
  \caption{The values of the scaling exponents of the characteristic length scale $L$ (black solid), the velocity $v$ (blue dash-dotted), the wiggliness $\mu$ (red dotted) and the correlation length $\xi$ (green dashed) for the growing wiggliness solutions given by Eq. (\ref{case4c}) and  Eq. (\ref{case4d}) as a function of the expansion rate $\lambda$.}
  \label{fig03}
\end{figure}

On the other hand, there is also a regime with solutions with growing wiggliness ($\gamma>0$). These require $\alpha=3\lambda/2$, implying the physical constraint $\lambda<2/3$, in other words expansion rates slower than that of the matter era. They further require $\beta<0$ (implying decaying velocities), $3\lambda/2-\beta+\gamma/2=1$, and $\beta+\gamma\ge0$. There are two possible branches of this solution, depending on the expansion rate. For very slow expansion rates we have
\be\label{case4c}
\begin{split}
    L &= \zeta_0t^{3\lambda/2}\\
    v &= v_0t^{-\lambda}\\
    \mu &= m_0t^{2-5\lambda}\\
    \xi &= \frac{kv_0}{2-4\lambda}t^{1-\lambda}\,,
\end{split}
\ee
subject to the constraints
\bq
\sqrt{m_0}\zeta_0&=&\frac{kv_0}{2-4\lambda}\\
\lambda&\le&\frac{1}{3}\,,
\eq
while for intermediate ones (including the radiation era, but not the matter era) we have
\be\label{case4d}
\begin{split}
    L &= \zeta_0t^{3\lambda/2}\\
    v &= \frac{t^{\lambda-2/3}}{\sqrt{3\lambda-1}m_0}\\
    \mu &= m_0t^{2/3-\lambda}\\
    \xi &= \frac{3}{2}kv_0t^{\lambda+1/3}\,,
\end{split}
\ee
subject to the constraints
\bq
\sqrt{m_0}\zeta_0&=&\frac{3}{2}kv_0\\
\frac{1}{3}<&\lambda&<\frac{2}{3}\,;
\eq
note that this solution matches that of Eq. (\ref{case4c}) for $\lambda=1/3$ and that of Eq. (\ref{case4b}) in the limit $\lambda\longrightarrow2/3$, except for a different normalization of the velocity in the latter. The behaviour of the scaling exponents in these solutions is depicted in Fig. \ref{fig03}. This illustrates the fact that a faster expansion rate leads to a slower growth of the wiggliness, This also highlights the special nature of the matter era case, $\lambda=2/3$, where one can have a full scaling solution, given by Eq. (\ref{case4b}). On the other hand, in the radiation era, with $\lambda=1/2$, we have
\be\label{case4e}
\begin{split}
    L &= \zeta_0t^{3/4}\\
    v &= \frac{\sqrt{2}}{m_0}t^{-1/6}\\
    \mu &=m_0t^{1/6}\\
    \xi &= \frac{3}{2}kv_0t^{5/6}\,.
\end{split}
\ee
It's also interesting to note that in the limit $\lambda\longrightarrow0$ we would have
\be\label{case4f}
\begin{split}
    L &= const.\\
    v &= const.\\
    \mu &=m_0t^{2}\\
    \xi &= \frac{kv_0}{2}t\,;
\end{split}
\ee
here the characteristic length scale would be constant (a consequence of energy conservation), as would the velocity, while the correlation length would scale linearly. The other peculiar case is $\lambda=1/3$, which leads to the fastest decaying velocities and the slowest growing correlation length. We note that these solutions could be tested in Nambu-Goto simulations, where energy loss terms can be switched off at will.

Finally, we note that in both of these growing wiggliness solutions the ratio of the total string energy density grows with respect to the background energy density as
\be
\frac{\rho}{\rho_{bckg}}\propto t^{2-3\lambda}\,,
\ee
while the ratio of the bare string energy to the background energy density grows as
\be
\begin{split}
    \frac{\rho_0}{\rho_{bckg}}&\propto t^{\lambda}\,,\qquad \lambda\le\frac{1}{3} \\
    \frac{\rho_0}{\rho_{bckg}}&\propto t^{4/3-2\lambda}\,,\qquad \frac{1}{3}<\lambda<\frac{2}{3}\,
\end{split}
\ee
respectively for the slow and intermediate expansion rate ranges. For the particular case of the radiation era, these two ratios should therefore be
\be
\begin{split}
    \frac{\rho_{rad}}{\rho_{bckg}}&\propto t^{1/2}\\
    \frac{\rho_{0,rad}}{\rho_{bckg}}&\propto t^{1/3}\,,
\end{split}
\ee
again it would be interesting to search for this solution in Nambu-Goto radiation era numerical simulations.

\subsection{With energy losses}

In this more realistic case, where both dynamical mechanisms are allowed, there are also three possible regimes. These are extensions of the ones in the previous sub-section, though the conditions to which each of them is subject change in interesting ways. Firstly, we have the canonical VOS model Nambu-Goto solution
\be\label{case6a}
\begin{split}
    L &= \zeta_0t\\
    v &= v_0\\
    \mu &= 1\\
    \xi &= \zeta_0t\,,
\end{split}
\ee
with the scaling parameters being given by
\bq
\zeta^2_{NG}&=&\frac{k(k+c)}{4\lambda(1-\lambda)}\\
v^2_{NG}&=&\frac{(1-\lambda)k}{\lambda(k+c)}\,;
\eq
these parameters are precisely the ones discussed in Eq. (\ref{scalsol}). In the $c=0$ limit we recover the previous solution given by Eq. (\ref{case4a}). The consistency condition relating the expansion rate and the VOS model parameters is $\lambda\ge 2k/(3k+c)$ if one imposes $v_0^2\le1/2$, or $\lambda>k/(2k+c)$ if one only requires $v_0^2<1$,

\subsubsection{Full scaling}

A full scaling solution, with a non-trivial constant wiggliness, $\mu_0>1$, also exists in principle, being given by
\be\label{case6b}
\begin{split}
    L &= \zeta_0t\\
    v &= v_0\\
    \mu &= m_0\\
    \xi &= \frac{k}{\lambda v_0(1+m_0^2)}t\,,
\end{split}
\ee
with the following normalization factors
\bq
\zeta_0^2&=&\frac{k\left(k+m_0^2c\left[1+\eta\left(1-m_0^{-1/2}\right)\right]\right)}{\lambda m_0(1+m_0^2)[\lambda+(2-3\lambda)m_0^2]}\\
v_0^2&=&\frac{k[\lambda+(2-3\lambda)m_0^2]}{\lambda(1+m_0^2)\left(k+m_0^2c\left[1+\eta\left(1-m_0^{-1/2}\right)\right]\right)}\,,
\eq
and the scaling value of the small-scale wiggliness being given, as a function of the expansion rate $\lambda$ and the VOS model parameters $c$, $k$, $\eta$ and $D$, by the relation
\bw
\be
[\lambda+m_0^2(2-3\lambda)]\left[(k+cD)\left(1-\frac{1}{m_0^2}\right)-c\eta\left(1-\frac{1}{\sqrt{m_0}}\right)\right]=\lambda\left(1-\frac{1}{m_0^2}\right) \left(k+m_0^2c\left[1+\eta\left(1-m_0^{-1/2}\right)\right]\right)\,.
\ee
\ew
Note that in principle this can exist for various expansion rates, and that in the particular case of $c=0$ and $\lambda=2/3$ one recovers the previous solution given in Eq. (\ref{case4b}). Indeed it is instructive to consider the two limiting cases, for large and small wiggliness. For $m_0\gg1$ we have
\be
\begin{split}
    L &= \sqrt{\frac{kc(1+\eta)}{\lambda(2-3\lambda)}}\frac{t}{m_0^{3/2}}\\
    v &= \sqrt{\frac{(2-3\lambda)k}{\lambda c(1+\eta)}}\frac{1}{m_0}\\
    \mu &= m_0\\
    \xi &= \sqrt{\frac{kc(1+\eta)}{\lambda(2-3\lambda)}}\frac{t}{m_0}\,,
\end{split}
\ee
and we confirm the expectation that in this case the density increases and the velocity and correlation length decrease with increasing wiggliness. This solution exists for a single expansion rate, given by
\be\label{fullscaling}
\lambda=\frac{2k_{eff}}{3k_{eff}+c_{eff}}\,,
\ee
where for convenience we have defined
\bq\label{effk}
k_{eff}&=&k+c(D-\eta)\\ 
c_{eff}&=&c(1+\eta)\,, \label{effc}
\eq
and again one trivially sees that the previous matter era solution is recovered if there are no energy losses. Note that these two definitions make physical sense: the presence of small-scale structure on the strings changes their typical curvature and enhances energy losses. From this it follows that for this solution to exist in the radiation era we require
\be\label{radscaling}
k_{eff}=c_{eff}\,,
\ee
or equivalently
\be
k=c(1+2\eta-D)\,.
\ee
This provides a possible calibration test for the model using numerical simulations: observation of scaling of small-scale structures in the radiation era would imply that Eq. (\ref{radscaling}) should hold.

In the opposite limit of small wiggliness we can Taylor-expand, defining $\mu\sim 1 + y$. and the scaling solution has the form
\be
\begin{split}
    L &= \zeta_0t\\
    v &= v_0\\
    \mu &= 1+y_0\\
    \xi &= \frac{k(1-y_0)}{2\lambda v_0}t\,,
\end{split}
\ee
with the following normalization factors, conveniently expressed in terms of first-order corrections to the Nambu-Goto solution
\bq
\zeta_0^2&=&\zeta_{NG}^2\left[1-\frac{4-5\lambda}{1-\lambda}y_0+\frac{c}{k+c}\left(2+\frac{1}{2}\eta\right)y_0\right]\\
v_0^2&=&v_{NG}^2\left[1+\frac{1-2\lambda}{1-\lambda}y_0-\frac{c}{k+c}\left(2+\frac{1}{2}\eta\right)y_0\right]\,,
\eq
together with
\be
y_0=\frac{\lambda(k+c)-(1-\lambda)[2(k+cD)-\eta c/2]}{(2-3\lambda)[2(k+cD)-\eta c/2]-\lambda c(2+\eta/2)}\,.
\ee

\begin{figure}
\centering
  \includegraphics[width=1.00\columnwidth]{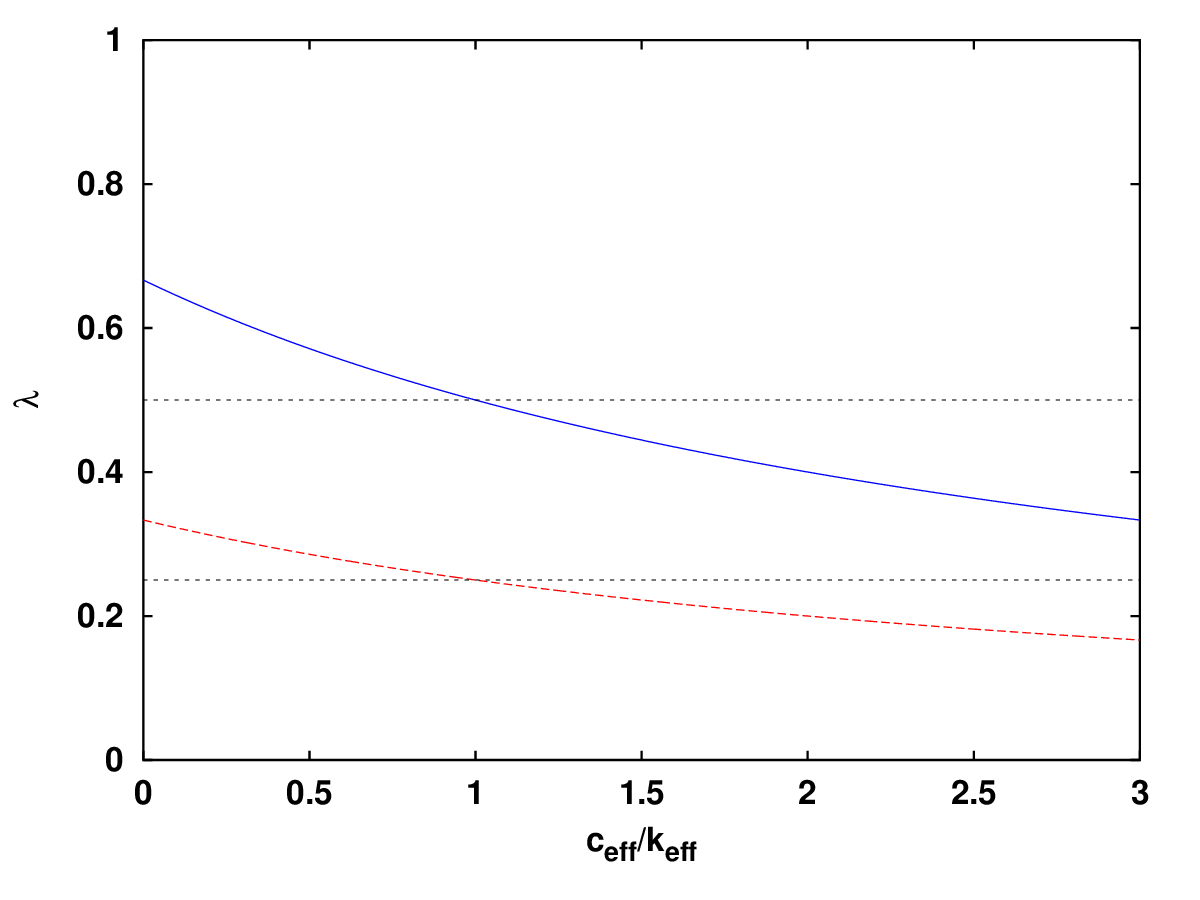}
  \caption{The behaviour of the expansion rate of the transition between slow and intermediate expansion regimes for growing wiggliness, given by Eq. (\ref{lambdamid}) (red dashed line), and the expansion rate of full scaling given by Eq. (\ref{fullscaling}) (blue solid line), as a function of the ratio $c_{eff}/k_{eff}$.}
  \label{fig04}
\end{figure}

\subsubsection{Growing wiggliness}

The regime with with growing wiggliness again contains two solutions. These are analogous to the ones in Eqs. (\ref{case4c}) and (\ref{case4d}) and still require $\beta<0$ (implying decaying velocities), $\beta+\gamma\ge0$, and $\alpha-\beta+\gamma/2=1$. However, it is no longer the case that $\alpha=3\lambda/2$. Instead we must have $\alpha>3\lambda/2$, which again makes sense: the addition of energy losses implies that the total energy of the network will decay faster. Moreover, both $\alpha$ and $\lambda$ now depend on the VOS model parameters, and there is a general relation
\be
\frac{v_0}{\sqrt{m_0}\zeta_0}=\frac{\lambda+\gamma}{k_{eff}}=\frac{2\alpha-3\lambda}{c_{eff}}\,,
\ee
with $k_{eff}$ and $c_{eff}$ as defined in Eq. (\ref{effk}) and Eq. (\ref{effc}) respectively.

For very slow expansion rates we have
\be\label{case6c}
\begin{split}
    L &= \zeta_0t^{1-\lambda-\gamma/2}\\
    v &= v_0t^{-\lambda}\\
    \mu &= m_0t^{\gamma}\\
    \xi &= \sqrt{m_0}\zeta_0t^{1-\lambda}\,,
\end{split}
\ee
with the following consistency conditions and restrictions
\bq
\gamma&=&\frac{(2-5\lambda)k_{eff}-\lambda c_{eff}}{k_{eff}+c_{eff}}\\
0<&\gamma&<2-5\lambda\\
\lambda&\le&\frac{k_{eff}}{3k_{eff}+c_{eff}}\,,
\eq
while for intermediate ones we have
\be\label{case6d}
\begin{split}
    L &= \zeta_0t^{1-3\gamma/2}\\
    v &= v_0t^{-\gamma}\\
    \mu &= m_0t^{\gamma}\\
    \xi &= \sqrt{m_0}\zeta_0t^{1-\gamma}\,,
\end{split}
\ee
with the following consistency conditions and restrictions
\bq
\gamma&=&\frac{(2-3\lambda)k_{eff}-\lambda c_{eff}}{3k_{eff}+c_{eff}}\\
\frac{v_0}{\sqrt{m_0}\zeta_0}&=&(\lambda-\gamma)\frac{v_0^2m_0^2}{k}\\
0<&\gamma&<min(\lambda,2-3\lambda)\\
\frac{k_{eff}}{3k_{eff}+c_{eff}}<&\lambda&<\frac{2k_{eff}}{3k_{eff}+c_{eff}}\,;
\eq
note that these solutions match those of Eqs. (\ref{case4c}) and (\ref{case4d}) in the limit $c\longrightarrow0$, and that the two solutions match for the expansion rate
\be\label{lambdamid}
\lambda=\frac{k_{eff}}{3k_{eff}+c_{eff}}\,,
\ee
which approaches $\lambda=1/3$ in the limit $c\longrightarrow0$. We depict the dependence of this expansion rate, as well as the expansion rate of full scaling given by Eq. (\ref{fullscaling}), on the ratio $c_{eff}/k_{eff}$, in Fig. \ref{fig04}.

In this case the ratios between the total string and background density and the bare string and background density are
\be
\begin{split}
    \frac{\rho_{slow}}{\rho_{bckg}}&\propto t^{2\lambda+\gamma}\\
    \frac{\rho_{0,slow}}{\rho_{bckg}}&\propto t^{2\lambda}\,
\end{split}
\ee
for slow expansion rates, and
\be
\begin{split}
    \frac{\rho_{int}}{\rho_{bckg}}&\propto t^{3\gamma}\\
    \frac{\rho_{0,int}}{\rho_{bckg}}&\propto t^{2\gamma}\,
\end{split}
\ee
for intermediate expansion rates.

Finally, it's interesting to consider the particular case of $k_{eff}=c_{eff}$, which leads to full scaling in the radiation era. In this case the scaling exponents depend only on the expansion rate, and in particular the one for the wiggliness parameter has the form
\be
\begin{split}
    \gamma_{slow}&=1-3\lambda\,,\qquad \lambda\le\frac{1}{4} \\
    \gamma_{int}&=\frac{1}{2}-\lambda\,,\qquad \frac{1}{4}<\lambda<\frac{1}{2}\,;
\end{split}
\ee
naturally in this case the growing wiggliness solutions only exist for expansion rates slower than the radiation era one, and the transition between slow and intermediate expansion rates now occurs at $\lambda=1/4$.

\section{\label{concl}Conclusions}

We have improved the physical interpretation of the wiggly extension of the VOS model \cite{PAP1,PAP2}, by studying its possible asymptotic scaling solutions of this model, and in particular how they depend on the expansion of the universe and the available energy loss or transfer mechanisms---e.g., the production of loops and wiggles. We have restricted ourselves to the case of a constant renormalization scale. This is clearly a simplifying assumption, but it is justified as a means to gain physical insight on the role of the other dynamical mechanisms, and the parameters that model them, in the evolution of the network. Relaxing this assumption, thus allowing for a running renormalization scale, is left for subsequent work.

Without expansion (i.e., in Minkowski space) the scaling solution is a trivial equilibrium solution in the absence of energy losses, while if these losses are allowed we find that asymptotically the correlation length $\xi$ always scales, while the characteristic length $L$ might scale linearly (with constant wiggliness) or more slowly than that (with growing wiggliness), depending on the balance between the amount of wiggles produced and removed from the network. These results are compatible with the observation of linear scaling in Minkowski space simulations \cite{Sakellariadou,FRAC}, 

In power law expanding universes there are three types of solutions, primarily depending on the expansion rate. For slow expansion rates one has growing wiggliness solutions, while for fast expansion rates we have the Nambu-Goto one (without wiggles). In between the two, for a single value of the expansion rate, there is a full scaling solution, with $L\propto\xi\propto t$ and constant velocity and wiggliness. Without energy losses this specific expansion rate corresponds to the matter era, while the addition of energy losses makes this transition expansion rate decrease---possibly reaching the radiation era for an appropriate value of the energy losses. These results are also compatible with previous Nambu-Goto simulation studies which indicate that full scaling of the network, including the wiggliness, is much more likely in the matter era than in the radiation era \cite{BB,AS,FRAC}. It is also interesting to note that these three types of solutions, depending on the expansion rate and with a transition expansion rate that depends on the amount of energy losses and corresponds to the matter era in their absence, also occurs for chiral superconducting strings \cite{Oliveira}.

The next step is to carry out a detailed comparison of these solution with numerical simulations. Some data from Nambu-Goto simulations already exists \cite{BB,AS,Sakellariadou,FRAC,Blanco}, and while our results are in qualitative agreement with these works, the data therein is not precise or complete enough to carry out a quantitative comparison. Our work therefore provides motivation for additional, higher resolution simulations. We note that Nambu-Goto simulations are useful for testing some of the new solutions we have presented, since in these simulations one can switch intercommuting and loop production on and off at will.

A similar comparison should also be done between these solutions and field theory simulations. Traditional CPU-based Abelian-Higgs simulations
\cite{ABELIAN,Stuckey,Hiramatsu} clearly lack the spatial resolution and dynamic range to study small-scale wiggliness, despite previous efforts \cite{Stuckey}. Recently a new generation GPU-accelerated Abelian-Higgs simulation code \cite{GPU1,GPU2} has emerged, enabling a detailed and statistically more robust calibration of the vOS model \cite{Correia1,Correia2}. This fast and efficient code will enable an independent analysis of the small-scale structures on cosmic string networks which, in addition to testing and calibrating some of the scaling solutions we have discussed, is also important as a comparison between Nambu-Goto and Abelian-Higgs simulations per se. Work along these lines is ongoing.

\begin{acknowledgments}
This work was financed by FEDER---Fundo Europeu de Desenvolvimento Regional funds through the COMPETE 2020---Operational Programme for Competitiveness and Internationalisation (POCI), and by Portuguese funds through FCT - Funda\c c\~ao para a Ci\^encia e a Tecnologia in the framework of the project POCI-01-0145-FEDER-028987 and PTDC/FIS-AST/28987/2017.

Discussions with Jos\'e Ricardo Correia, Filipe Costa, Patrick Peter, Ivan Rybak, Paul Shellard and Jos\'e Pedro Vieira are gratefully acknowledged.
\end{acknowledgments}

\bibliography{wiggly}

\begin{thebibliography}{34}%
\makeatletter
\providecommand \@ifxundefined [1]{%
 \@ifx{#1\undefined}
}%
\providecommand \@ifnum [1]{%
 \ifnum #1\expandafter \@firstoftwo
 \else \expandafter \@secondoftwo
 \fi
}%
\providecommand \@ifx [1]{%
 \ifx #1\expandafter \@firstoftwo
 \else \expandafter \@secondoftwo
 \fi
}%
\providecommand \natexlab [1]{#1}%
\providecommand \enquote  [1]{``#1''}%
\providecommand \bibnamefont  [1]{#1}%
\providecommand \bibfnamefont [1]{#1}%
\providecommand \citenamefont [1]{#1}%
\providecommand \href@noop [0]{\@secondoftwo}%
\providecommand \href [0]{\begingroup \@sanitize@url \@href}%
\providecommand \@href[1]{\@@startlink{#1}\@@href}%
\providecommand \@@href[1]{\endgroup#1\@@endlink}%
\providecommand \@sanitize@url [0]{\catcode `\\12\catcode `\$12\catcode
  `\&12\catcode `\#12\catcode `\^12\catcode `\_12\catcode `\%12\relax}%
\providecommand \@@startlink[1]{}%
\providecommand \@@endlink[0]{}%
\providecommand \url  [0]{\begingroup\@sanitize@url \@url }%
\providecommand \@url [1]{\endgroup\@href {#1}{\urlprefix }}%
\providecommand \urlprefix  [0]{URL }%
\providecommand \Eprint [0]{\href }%
\providecommand \doibase [0]{http://dx.doi.org/}%
\providecommand \selectlanguage [0]{\@gobble}%
\providecommand \bibinfo  [0]{\@secondoftwo}%
\providecommand \bibfield  [0]{\@secondoftwo}%
\providecommand \translation [1]{[#1]}%
\providecommand \BibitemOpen [0]{}%
\providecommand \bibitemStop [0]{}%
\providecommand \bibitemNoStop [0]{.\EOS\space}%
\providecommand \EOS [0]{\spacefactor3000\relax}%
\providecommand \BibitemShut  [1]{\csname bibitem#1\endcsname}%
\let\auto@bib@innerbib\@empty
\bibitem [{\citenamefont {Kibble}(1976)}]{Kibble76}%
  \BibitemOpen
  \bibfield  {author} {\bibinfo {author} {\bibfnamefont {T.~W.~B.}\
  \bibnamefont {Kibble}},\ }\href {\doibase 10.1088/0305-4470/9/8/029}
  {\bibfield  {journal} {\bibinfo  {journal} {J. Phys. A}\ }\textbf {\bibinfo
  {volume} {9}},\ \bibinfo {pages} {1387} (\bibinfo {year} {1976})}\BibitemShut
  {NoStop}%
\bibitem [{\citenamefont {Vilenkin}\ and\ \citenamefont
  {Shellard}(1994)}]{VSH}%
  \BibitemOpen
  \bibfield  {author} {\bibinfo {author} {\bibfnamefont {A.}~\bibnamefont
  {Vilenkin}}\ and\ \bibinfo {author} {\bibfnamefont {E.~P.~S.}\ \bibnamefont
  {Shellard}},\ }\href@noop {} {\emph {\bibinfo {title} {Cosmic Strings and
  other Topological Defects}}}\ (\bibinfo  {publisher} {Cambridge University
  Press},\ \bibinfo {address} {Cambridge, U.K.},\ \bibinfo {year}
  {1994})\BibitemShut {NoStop}%
\bibitem [{\citenamefont {Bennett}\ and\ \citenamefont {Bouchet}(1990)}]{BB}%
  \BibitemOpen
  \bibfield  {author} {\bibinfo {author} {\bibfnamefont {D.~P.}\ \bibnamefont
  {Bennett}}\ and\ \bibinfo {author} {\bibfnamefont {F.~R.}\ \bibnamefont
  {Bouchet}},\ }\href@noop {} {\bibfield  {journal} {\bibinfo  {journal} {Phys.
  Rev.}\ }\textbf {\bibinfo {volume} {D41}},\ \bibinfo {pages} {2408} (\bibinfo
  {year} {1990})}\BibitemShut {NoStop}%
\bibitem [{\citenamefont {Allen}\ and\ \citenamefont {Shellard}(1990)}]{AS}%
  \BibitemOpen
  \bibfield  {author} {\bibinfo {author} {\bibfnamefont {B.}~\bibnamefont
  {Allen}}\ and\ \bibinfo {author} {\bibfnamefont {E.~P.~S.}\ \bibnamefont
  {Shellard}},\ }\href@noop {} {\bibfield  {journal} {\bibinfo  {journal}
  {Phys. Rev. Lett.}\ }\textbf {\bibinfo {volume} {64}},\ \bibinfo {pages}
  {119} (\bibinfo {year} {1990})}\BibitemShut {NoStop}%
\bibitem [{\citenamefont {Moore}\ \emph {et~al.}(2002)\citenamefont {Moore},
  \citenamefont {Shellard},\ and\ \citenamefont {Martins}}]{ABELIAN}%
  \BibitemOpen
  \bibfield  {author} {\bibinfo {author} {\bibfnamefont {J.~N.}\ \bibnamefont
  {Moore}}, \bibinfo {author} {\bibfnamefont {E.~P.~S.}\ \bibnamefont
  {Shellard}}, \ and\ \bibinfo {author} {\bibfnamefont {C.~J. A.~P.}\
  \bibnamefont {Martins}},\ }\href@noop {} {\bibfield  {journal} {\bibinfo
  {journal} {Phys. Rev.}\ }\textbf {\bibinfo {volume} {D65}},\ \bibinfo {pages}
  {023503} (\bibinfo {year} {2002})},\ \Eprint
  {http://arxiv.org/abs/hep-ph/0107171} {hep-ph/0107171} \BibitemShut {NoStop}%
\bibitem [{\citenamefont {Martins}\ and\ \citenamefont
  {Shellard}(2006)}]{FRAC}%
  \BibitemOpen
  \bibfield  {author} {\bibinfo {author} {\bibfnamefont {C.~J. A.~P.}\
  \bibnamefont {Martins}}\ and\ \bibinfo {author} {\bibfnamefont {E.~P.~S.}\
  \bibnamefont {Shellard}},\ }\href@noop {} {\bibfield  {journal} {\bibinfo
  {journal} {Phys. Rev.}\ }\textbf {\bibinfo {volume} {D73}},\ \bibinfo {pages}
  {043515} (\bibinfo {year} {2006})},\ \Eprint
  {http://arxiv.org/abs/astro-ph/0511792} {astro-ph/0511792} \BibitemShut
  {NoStop}%
\bibitem [{\citenamefont {Ringeval}\ \emph {et~al.}(2007)\citenamefont
  {Ringeval}, \citenamefont {Sakellariadou},\ and\ \citenamefont
  {Bouchet}}]{RSB}%
  \BibitemOpen
  \bibfield  {author} {\bibinfo {author} {\bibfnamefont {C.}~\bibnamefont
  {Ringeval}}, \bibinfo {author} {\bibfnamefont {M.}~\bibnamefont
  {Sakellariadou}}, \ and\ \bibinfo {author} {\bibfnamefont {F.}~\bibnamefont
  {Bouchet}},\ }\href@noop {} {\bibfield  {journal} {\bibinfo  {journal}
  {JCAP}\ }\textbf {\bibinfo {volume} {0702}},\ \bibinfo {pages} {023}
  (\bibinfo {year} {2007})},\ \Eprint {http://arxiv.org/abs/astro-ph/0511646}
  {astro-ph/0511646} \BibitemShut {NoStop}%
\bibitem [{\citenamefont {Olum}\ and\ \citenamefont {Vanchurin}(2007)}]{VVO}%
  \BibitemOpen
  \bibfield  {author} {\bibinfo {author} {\bibfnamefont {K.~D.}\ \bibnamefont
  {Olum}}\ and\ \bibinfo {author} {\bibfnamefont {V.}~\bibnamefont
  {Vanchurin}},\ }\href@noop {} {\bibfield  {journal} {\bibinfo  {journal}
  {Phys. Rev.}\ }\textbf {\bibinfo {volume} {D75}},\ \bibinfo {pages} {063521}
  (\bibinfo {year} {2007})},\ \Eprint {http://arxiv.org/abs/astro-ph/0610419}
  {astro-ph/0610419} \BibitemShut {NoStop}%
\bibitem [{\citenamefont {Hindmarsh}\ \emph {et~al.}(2009)\citenamefont
  {Hindmarsh}, \citenamefont {Stuckey},\ and\ \citenamefont {Bevis}}]{Stuckey}%
  \BibitemOpen
  \bibfield  {author} {\bibinfo {author} {\bibfnamefont {M.}~\bibnamefont
  {Hindmarsh}}, \bibinfo {author} {\bibfnamefont {S.}~\bibnamefont {Stuckey}},
  \ and\ \bibinfo {author} {\bibfnamefont {N.}~\bibnamefont {Bevis}},\ }\href
  {\doibase 10.1103/PhysRevD.79.123504} {\bibfield  {journal} {\bibinfo
  {journal} {Phys. Rev.}\ }\textbf {\bibinfo {volume} {D79}},\ \bibinfo {pages}
  {123504} (\bibinfo {year} {2009})},\ \Eprint {http://arxiv.org/abs/0812.1929}
  {arXiv:0812.1929 [hep-th]} \BibitemShut {NoStop}%
\bibitem [{\citenamefont {Blanco-Pillado}\ \emph {et~al.}(2011)\citenamefont
  {Blanco-Pillado}, \citenamefont {Olum},\ and\ \citenamefont
  {Shlaer}}]{Blanco}%
  \BibitemOpen
  \bibfield  {author} {\bibinfo {author} {\bibfnamefont {J.~J.}\ \bibnamefont
  {Blanco-Pillado}}, \bibinfo {author} {\bibfnamefont {K.~D.}\ \bibnamefont
  {Olum}}, \ and\ \bibinfo {author} {\bibfnamefont {B.}~\bibnamefont
  {Shlaer}},\ }\href {\doibase 10.1103/PhysRevD.83.083514} {\bibfield
  {journal} {\bibinfo  {journal} {Phys. Rev.}\ }\textbf {\bibinfo {volume}
  {D83}},\ \bibinfo {pages} {083514} (\bibinfo {year} {2011})},\ \Eprint
  {http://arxiv.org/abs/1101.5173} {arXiv:1101.5173 [astro-ph.CO]} \BibitemShut
  {NoStop}%
\bibitem [{\citenamefont {Hiramatsu}\ \emph {et~al.}(2013)\citenamefont
  {Hiramatsu}, \citenamefont {Sendouda}, \citenamefont {Takahashi},
  \citenamefont {Yamauchi},\ and\ \citenamefont {Yoo}}]{Hiramatsu}%
  \BibitemOpen
  \bibfield  {author} {\bibinfo {author} {\bibfnamefont {T.}~\bibnamefont
  {Hiramatsu}}, \bibinfo {author} {\bibfnamefont {Y.}~\bibnamefont {Sendouda}},
  \bibinfo {author} {\bibfnamefont {K.}~\bibnamefont {Takahashi}}, \bibinfo
  {author} {\bibfnamefont {D.}~\bibnamefont {Yamauchi}}, \ and\ \bibinfo
  {author} {\bibfnamefont {C.-M.}\ \bibnamefont {Yoo}},\ }\href {\doibase
  10.1103/PhysRevD.88.085021} {\bibfield  {journal} {\bibinfo  {journal} {Phys.
  Rev.}\ }\textbf {\bibinfo {volume} {D88}},\ \bibinfo {pages} {085021}
  (\bibinfo {year} {2013})},\ \Eprint {http://arxiv.org/abs/1307.0308}
  {arXiv:1307.0308 [astro-ph.CO]} \BibitemShut {NoStop}%
\bibitem [{\citenamefont {Correia}\ and\ \citenamefont
  {Martins}(2019)}]{Correia1}%
  \BibitemOpen
  \bibfield  {author} {\bibinfo {author} {\bibfnamefont {J.}~\bibnamefont
  {Correia}}\ and\ \bibinfo {author} {\bibfnamefont {C.}~\bibnamefont
  {Martins}},\ }\href {\doibase 10.1103/PhysRevD.100.103517} {\bibfield
  {journal} {\bibinfo  {journal} {Phys. Rev. D}\ }\textbf {\bibinfo {volume}
  {100}},\ \bibinfo {pages} {103517} (\bibinfo {year} {2019})},\ \Eprint
  {http://arxiv.org/abs/1911.03163} {arXiv:1911.03163 [astro-ph.CO]}
  \BibitemShut {NoStop}%
\bibitem [{\citenamefont {Correia}\ and\ \citenamefont
  {Martins}(2020{\natexlab{a}})}]{Correia2}%
  \BibitemOpen
  \bibfield  {author} {\bibinfo {author} {\bibfnamefont {J.}~\bibnamefont
  {Correia}}\ and\ \bibinfo {author} {\bibfnamefont {C.}~\bibnamefont
  {Martins}},\ }\href {\doibase 10.1103/PhysRevD.102.043503} {\bibfield
  {journal} {\bibinfo  {journal} {Phys. Rev. D}\ }\textbf {\bibinfo {volume}
  {102}},\ \bibinfo {pages} {043503} (\bibinfo {year} {2020}{\natexlab{a}})},\
  \Eprint {http://arxiv.org/abs/2007.12008} {arXiv:2007.12008 [astro-ph.CO]}
  \BibitemShut {NoStop}%
\bibitem [{\citenamefont {Correia}\ and\ \citenamefont
  {Martins}(2020{\natexlab{b}})}]{GPU1}%
  \BibitemOpen
  \bibfield  {author} {\bibinfo {author} {\bibfnamefont {J.~R. C. C.~C.}\
  \bibnamefont {Correia}}\ and\ \bibinfo {author} {\bibfnamefont {C.~J. A.~P.}\
  \bibnamefont {Martins}},\ }\href {\doibase 10.1016/j.ascom.2020.100388}
  {\bibfield  {journal} {\bibinfo  {journal} {Astron. Comput.}\ }\textbf
  {\bibinfo {volume} {32}},\ \bibinfo {pages} {100388} (\bibinfo {year}
  {2020}{\natexlab{b}})},\ \Eprint {http://arxiv.org/abs/1809.00995}
  {arXiv:1809.00995 [physics.comp-ph]} \BibitemShut {NoStop}%
\bibitem [{\citenamefont {Correia}\ and\ \citenamefont {Martins}(2021)}]{GPU2}%
  \BibitemOpen
  \bibfield  {author} {\bibinfo {author} {\bibfnamefont {J.~R. C. C.~C.}\
  \bibnamefont {Correia}}\ and\ \bibinfo {author} {\bibfnamefont {C.~J. A.~P.}\
  \bibnamefont {Martins}},\ }\href {\doibase 10.1016/j.ascom.2020.100438}
  {\bibfield  {journal} {\bibinfo  {journal} {Astron. Comput.}\ }\textbf
  {\bibinfo {volume} {34}},\ \bibinfo {pages} {100438} (\bibinfo {year}
  {2021})},\ \Eprint {http://arxiv.org/abs/2005.14454} {arXiv:2005.14454
  [physics.comp-ph]} \BibitemShut {NoStop}%
\bibitem [{\citenamefont {Martins}\ and\ \citenamefont {Shellard}(1996)}]{MS2}%
  \BibitemOpen
  \bibfield  {author} {\bibinfo {author} {\bibfnamefont {C.~J. A.~P.}\
  \bibnamefont {Martins}}\ and\ \bibinfo {author} {\bibfnamefont {E.~P.~S.}\
  \bibnamefont {Shellard}},\ }\href@noop {} {\bibfield  {journal} {\bibinfo
  {journal} {Phys. Rev.}\ }\textbf {\bibinfo {volume} {D54}},\ \bibinfo {pages}
  {2535} (\bibinfo {year} {1996})},\ \Eprint
  {http://arXiv.org/abs/hep-ph/9602271} {hep-ph/9602271} \BibitemShut {NoStop}%
\bibitem [{\citenamefont {Martins}\ and\ \citenamefont {Shellard}(2002)}]{MS3}%
  \BibitemOpen
  \bibfield  {author} {\bibinfo {author} {\bibfnamefont {C.~J. A.~P.}\
  \bibnamefont {Martins}}\ and\ \bibinfo {author} {\bibfnamefont {E.~P.~S.}\
  \bibnamefont {Shellard}},\ }\href@noop {} {\bibfield  {journal} {\bibinfo
  {journal} {Phys. Rev.}\ }\textbf {\bibinfo {volume} {D65}},\ \bibinfo {pages}
  {043514} (\bibinfo {year} {2002})},\ \Eprint
  {http://arXiv.org/abs/hep-ph/0003298} {hep-ph/0003298} \BibitemShut {NoStop}%
\bibitem [{\citenamefont {Martins}\ \emph {et~al.}(2004)\citenamefont
  {Martins}, \citenamefont {Moore},\ and\ \citenamefont {Shellard}}]{MS4}%
  \BibitemOpen
  \bibfield  {author} {\bibinfo {author} {\bibfnamefont {C.~J. A.~P.}\
  \bibnamefont {Martins}}, \bibinfo {author} {\bibfnamefont {J.~N.}\
  \bibnamefont {Moore}}, \ and\ \bibinfo {author} {\bibfnamefont {E.~P.~S.}\
  \bibnamefont {Shellard}},\ }\href@noop {} {\bibfield  {journal} {\bibinfo
  {journal} {Phys. Rev. Lett.}\ }\textbf {\bibinfo {volume} {92}},\ \bibinfo
  {pages} {251601} (\bibinfo {year} {2004})},\ \Eprint
  {http://arxiv.org/abs/hep-ph/0310255} {hep-ph/0310255} \BibitemShut {NoStop}%
\bibitem [{\citenamefont {Martins}(2016)}]{Book}%
  \BibitemOpen
  \bibfield  {author} {\bibinfo {author} {\bibfnamefont {C.~J. A.~P.}\
  \bibnamefont {Martins}},\ }\href@noop {} {\emph {\bibinfo {title} {Defect
  Evolution in Cosmology and Condensed Matter: Quantitative Analysis with the
  Velocity-Dependent One-Scale Model}}}\ (\bibinfo  {publisher} {Springer},\
  \bibinfo {year} {2016})\BibitemShut {NoStop}%
\bibitem [{\citenamefont {Austin}\ \emph {et~al.}(1993)\citenamefont {Austin},
  \citenamefont {Copeland},\ and\ \citenamefont {Kibble}}]{ACK}%
  \BibitemOpen
  \bibfield  {author} {\bibinfo {author} {\bibfnamefont {D.}~\bibnamefont
  {Austin}}, \bibinfo {author} {\bibfnamefont {E.~J.}\ \bibnamefont
  {Copeland}}, \ and\ \bibinfo {author} {\bibfnamefont {T.~W.~B.}\ \bibnamefont
  {Kibble}},\ }\href@noop {} {\bibfield  {journal} {\bibinfo  {journal} {Phys.
  Rev.}\ }\textbf {\bibinfo {volume} {D48}},\ \bibinfo {pages} {5594} (\bibinfo
  {year} {1993})},\ \Eprint {http://arxiv.org/abs/hep-ph/9307325}
  {hep-ph/9307325} \BibitemShut {NoStop}%
\bibitem [{\citenamefont {Polchinski}\ and\ \citenamefont
  {Rocha}(2007)}]{POLR}%
  \BibitemOpen
  \bibfield  {author} {\bibinfo {author} {\bibfnamefont {J.}~\bibnamefont
  {Polchinski}}\ and\ \bibinfo {author} {\bibfnamefont {J.~V.}\ \bibnamefont
  {Rocha}},\ }\href@noop {} {\bibfield  {journal} {\bibinfo  {journal} {Phys.
  Rev.}\ }\textbf {\bibinfo {volume} {D75}},\ \bibinfo {pages} {123503}
  (\bibinfo {year} {2007})},\ \Eprint {http://arxiv.org/abs/gr-qc/0702055}
  {gr-qc/0702055} \BibitemShut {NoStop}%
\bibitem [{\citenamefont {Martins}\ \emph {et~al.}(2014)\citenamefont
  {Martins}, \citenamefont {Shellard},\ and\ \citenamefont {Vieira}}]{PAP1}%
  \BibitemOpen
  \bibfield  {author} {\bibinfo {author} {\bibfnamefont {C.~J. A.~P.}\
  \bibnamefont {Martins}}, \bibinfo {author} {\bibfnamefont {E.~P.~S.}\
  \bibnamefont {Shellard}}, \ and\ \bibinfo {author} {\bibfnamefont {J.~P.~P.}\
  \bibnamefont {Vieira}},\ }\href {\doibase 10.1103/PhysRevD.90.043518}
  {\bibfield  {journal} {\bibinfo  {journal} {Phys. Rev. D}\ }\textbf {\bibinfo
  {volume} {90}},\ \bibinfo {pages} {043518} (\bibinfo {year}
  {2014})}\BibitemShut {NoStop}%
\bibitem [{\citenamefont {Vieira}\ \emph {et~al.}(2016)\citenamefont {Vieira},
  \citenamefont {Martins},\ and\ \citenamefont {Shellard}}]{PAP2}%
  \BibitemOpen
  \bibfield  {author} {\bibinfo {author} {\bibfnamefont {J.}~\bibnamefont
  {Vieira}}, \bibinfo {author} {\bibfnamefont {C.}~\bibnamefont {Martins}}, \
  and\ \bibinfo {author} {\bibfnamefont {E.}~\bibnamefont {Shellard}},\ }\href
  {\doibase 10.1103/PhysRevD.94.096005} {\bibfield  {journal} {\bibinfo
  {journal} {Phys. Rev. D}\ }\textbf {\bibinfo {volume} {94}},\ \bibinfo
  {pages} {096005} (\bibinfo {year} {2016})},\ \bibinfo {note} {[Erratum:
  Phys.Rev.D 94, 099907 (2016)]},\ \Eprint {http://arxiv.org/abs/1611.06103}
  {arXiv:1611.06103 [astro-ph.CO]} \BibitemShut {NoStop}%
\bibitem [{\citenamefont {Martins}\ \emph {et~al.}(2020)\citenamefont
  {Martins}, \citenamefont {Peter}, \citenamefont {Rybak},\ and\ \citenamefont
  {Shellard}}]{Currents}%
  \BibitemOpen
  \bibfield  {author} {\bibinfo {author} {\bibfnamefont {C.~J. A.~P.}\
  \bibnamefont {Martins}}, \bibinfo {author} {\bibfnamefont {P.}~\bibnamefont
  {Peter}}, \bibinfo {author} {\bibfnamefont {I.~Y.}\ \bibnamefont {Rybak}}, \
  and\ \bibinfo {author} {\bibfnamefont {E.~P.~S.}\ \bibnamefont {Shellard}},\
  }\href@noop {} {\  (\bibinfo {year} {2020})},\ \Eprint
  {http://arxiv.org/abs/2011.09700} {arXiv:2011.09700 [astro-ph.CO]}
  \BibitemShut {NoStop}%
\bibitem [{\citenamefont {Kibble}(1985)}]{KIB}%
  \BibitemOpen
  \bibfield  {author} {\bibinfo {author} {\bibfnamefont {T.~W.~B.}\
  \bibnamefont {Kibble}},\ }\href@noop {} {\bibfield  {journal} {\bibinfo
  {journal} {Nucl. Phys.}\ }\textbf {\bibinfo {volume} {B252}},\ \bibinfo
  {pages} {227} (\bibinfo {year} {1985})}\BibitemShut {NoStop}%
\bibitem [{\citenamefont {Bennett}(1986)}]{BMOD}%
  \BibitemOpen
  \bibfield  {author} {\bibinfo {author} {\bibfnamefont {D.~P.}\ \bibnamefont
  {Bennett}},\ }\href@noop {} {\bibfield  {journal} {\bibinfo  {journal} {Phys.
  Rev.}\ }\textbf {\bibinfo {volume} {D33}},\ \bibinfo {pages} {872} (\bibinfo
  {year} {1986})}\BibitemShut {NoStop}%
\bibitem [{\citenamefont {Azevedo}\ and\ \citenamefont
  {Martins}(2017)}]{Azevedo}%
  \BibitemOpen
  \bibfield  {author} {\bibinfo {author} {\bibfnamefont {R.~P.~L.}\
  \bibnamefont {Azevedo}}\ and\ \bibinfo {author} {\bibfnamefont {C.~J. A.~P.}\
  \bibnamefont {Martins}},\ }\href {\doibase 10.1103/PhysRevD.95.043537}
  {\bibfield  {journal} {\bibinfo  {journal} {Phys. Rev. D}\ }\textbf {\bibinfo
  {volume} {95}},\ \bibinfo {pages} {043537} (\bibinfo {year} {2017})},\
  \Eprint {http://arxiv.org/abs/1702.08453} {arXiv:1702.08453 [astro-ph.CO]}
  \BibitemShut {NoStop}%
\bibitem [{\citenamefont {Carter}(1995)}]{CARTERA}%
  \BibitemOpen
  \bibfield  {author} {\bibinfo {author} {\bibfnamefont {B.}~\bibnamefont
  {Carter}},\ }\href@noop {} {\bibfield  {journal} {\bibinfo  {journal} {Phys.
  Rev. Lett.}\ }\textbf {\bibinfo {volume} {74}},\ \bibinfo {pages} {3098}
  (\bibinfo {year} {1995})},\ \Eprint {http://arxiv.org/abs/hep-th/9411231}
  {hep-th/9411231} \BibitemShut {NoStop}%
\bibitem [{\citenamefont {Carter}(1997)}]{CARTERB}%
  \BibitemOpen
  \bibfield  {author} {\bibinfo {author} {\bibfnamefont {B.}~\bibnamefont
  {Carter}},\ }in\ \href@noop {} {\emph {\bibinfo {booktitle} {{2nd Mexican
  School on Gravitation and Mathematical Physics}}}}\ (\bibinfo {year} {1997})\
  \Eprint {http://arxiv.org/abs/hep-th/9705172} {arXiv:hep-th/9705172}
  \BibitemShut {NoStop}%
\bibitem [{\citenamefont {Takayasu}(1990)}]{TAKAYASU}%
  \BibitemOpen
  \bibfield  {author} {\bibinfo {author} {\bibfnamefont {H.}~\bibnamefont
  {Takayasu}},\ }\href@noop {} {\emph {\bibinfo {title} {Fractals in the
  Physical Sciences}}}\ (\bibinfo  {publisher} {Manchester University Press},\
  \bibinfo {address} {Manchester, U. K.},\ \bibinfo {year} {1990})\BibitemShut
  {NoStop}%
\bibitem [{\citenamefont {Sakellariadou}\ and\ \citenamefont
  {Vilenkin}(1990)}]{Sakellariadou}%
  \BibitemOpen
  \bibfield  {author} {\bibinfo {author} {\bibfnamefont {M.}~\bibnamefont
  {Sakellariadou}}\ and\ \bibinfo {author} {\bibfnamefont {A.}~\bibnamefont
  {Vilenkin}},\ }\href {\doibase 10.1103/PhysRevD.42.349} {\bibfield  {journal}
  {\bibinfo  {journal} {Phys. Rev. D}\ }\textbf {\bibinfo {volume} {42}},\
  \bibinfo {pages} {349} (\bibinfo {year} {1990})}\BibitemShut {NoStop}%
\bibitem [{\citenamefont {Martins}\ and\ \citenamefont
  {Cabral}(2016)}]{Cabral}%
  \BibitemOpen
  \bibfield  {author} {\bibinfo {author} {\bibfnamefont {C.~J. A.~P.}\
  \bibnamefont {Martins}}\ and\ \bibinfo {author} {\bibfnamefont {M.~M. P.
  V.~P.}\ \bibnamefont {Cabral}},\ }\href {\doibase 10.1103/PhysRevD.93.043542}
  {\bibfield  {journal} {\bibinfo  {journal} {Phys. Rev. D}\ }\textbf {\bibinfo
  {volume} {93}},\ \bibinfo {pages} {043542} (\bibinfo {year} {2016})},\
  \bibinfo {note} {[Addendum: Phys.Rev.D 93, 069902 (2016)]},\ \Eprint
  {http://arxiv.org/abs/1602.08083} {arXiv:1602.08083 [hep-ph]} \BibitemShut
  {NoStop}%
\bibitem [{Note1()}]{Note1}%
  \BibitemOpen
  \bibinfo {note} {We thank Jos\'e Pedro Vieira for bringing this
  interpretation to our attention.}\BibitemShut {Stop}%
\bibitem [{\citenamefont {Oliveira}\ \emph {et~al.}(2012)\citenamefont
  {Oliveira}, \citenamefont {Avgoustidis},\ and\ \citenamefont
  {Martins}}]{Oliveira}%
  \BibitemOpen
  \bibfield  {author} {\bibinfo {author} {\bibfnamefont {M.~F.}\ \bibnamefont
  {Oliveira}}, \bibinfo {author} {\bibfnamefont {A.}~\bibnamefont
  {Avgoustidis}}, \ and\ \bibinfo {author} {\bibfnamefont {C.~J. A.~P.}\
  \bibnamefont {Martins}},\ }\href {\doibase 10.1103/PhysRevD.85.083515}
  {\bibfield  {journal} {\bibinfo  {journal} {Phys. Rev. D}\ }\textbf {\bibinfo
  {volume} {85}},\ \bibinfo {pages} {083515} (\bibinfo {year} {2012})},\
  \Eprint {http://arxiv.org/abs/1201.5064} {arXiv:1201.5064 [hep-ph]}
  \BibitemShut {NoStop}%
\end{thebibliography}%
\end{document}